\documentclass{aa}             
\usepackage{amssymb,graphicx}  

\begin{document}

\thesaurus{08.14.2; 08.13.1; 08.12.2; 08.23.1}
\title{On the ultimate fate of AM Her stars}
\author{F. Meyer, E. Meyer-Hofmeister}

\institute{Max-Planck-Institut f\"ur Astrophysik, Karl
Schwarzschildstr.1, D-85740 Garching, Germany}

\offprints{Emmi Meyer-Hofmeister}

\date{Received:/ Accepted:}

\maketitle

\begin{abstract}

We suggest, that the magnetic field of the white dwarf in AM Her systems
loses coupling to the secondary star when the latter becomes
non-magnetic at the transition from a late main sequence star to a 
cool degenerate brown
dwarf. This leads to spin-up of the primary white dwarf. After
synchronous  rotation is lost the systems do not appear
as AM Her stars anymore. We discuss the further evolution of such systems.


\end{abstract}

\section{Introduction}

AM Her stars (polars) are magnetic cataclysmic variables where
the white dwarf primary and the Roche lobe filling low-mass secondary
star rotate
synchronously with the orbit. During secular evolution the
secondary star loses mass to the primary. The matter flow is
channeled by the magnetic field and accretes via
accretion columns near magnetic poles (King 1995). 
The magnetic coupling of white dwarf and
secondary star causes the synchronous rotation.

Secular evolution is the same for magnetic and non-magnetic systems.
It is commonly accepted that below the period gap
gravitational radiation causes the loss of angular momentum from the
binary system and the decrease of the orbital period. For stars of
mass below about 0.2$M_\odot$ the effective temperatur drops,
for around 0.06$M_\odot$ a transition to a cooling brown dwarf occurs.
The mean density reaches a maximum and
decreases again when the star becomes degenerate. Correspondingly 
Roche lobe and orbital period increase again
(Paczy\'nsky \& Sienkiewicz 1981, Rappaport et al. 1982, Ritter 1986).
For recent stellar modelling see 
Baraffe et al. (1998), Allard et al. (1996) and Allard et al. (1997). 

But stellar structure calculations consistently yielded a minimum
period of 70 minutes (most recently Kolb \& Baraffe 1999), significantly
lower than the observed cut off in the orbital period distribution of
cataclysmic variables (CVs) (Ritter \& Kolb 1998), near 80
minutes. To explain the missing (non-magnetic) dwarf nova systems
with periods between 80 and 70 minutes a selection effect (rare or no
outbursts) has been invoked. For the (magnetic) AM Her systems the shortest
period observed is 77.8 minutes (RXJ0132.7-6554, Burwitz et al. 1997).
Kolb \& Baraffe (1999) found that the
theoretical period minimum can be raised to 
80 minutes if the braking would be four times the
gravitational wave value. This might point to a residual braking from
the secondary star's magnetic field. We note that Patterson (1998)
pointed out a discrepancy between the predictions of CV evolution and
observations: far too few CVs are observed which have evolved past
period minimum.

We here propose the disappearance of AM Her systems after the period
turning point.
We argue that the magnetic field of the white dwarf loses coupling
to the secondary star when the secondary becomes non-magnetic at the
transition from a late main-sequence star to a brown dwarf. This leads to
spin-up of the primary white dwarf and the binary does no longer appear as
an AM Her system. Such a spin-up could lead to ejection of mass
and magnetic flux in a ``propeller phase'', a process which would
finally stop further spin-up. It is an interesting question how the stars
will appear after this metamorphosis.

In Sect. 2 we discuss the mechanisms of magnetic coupling and the
loss of coupling with the transition of the secondary to the brown dwarf state.
In Sect. 3 and 4 we consider the spin-up phase and a possible
propeller phase. In Sect. 5 the further evolution is
discussed.

\section{Magnetic coupling}

Various mechanisms have been suggested for the magnetic
coupling. Generation of a magnetic torque by a small degree of
asynchronism, dipole-dipole interaction, and conductive connection of
field lines between the two stars were considered. Even if no
secondary star magnetic field would exist convective mixing-in of
primary fields has been appealed to.
(Campbell 1985,1986, 1989, Lamb 1985, Lamb \& Melia 1988).
We discuss in the following why these conditions for coupling disappear
in the late state of secular binary evolution.

\subsection{Disappearance of the secondary's magnetic field}

In the transition from a low-mass main-sequence star to a
brown dwarf the overadiabatic structure and convection disappear. No
convection-based dynamo exists and no containment of captured flux in
a convective zone (Meyer 1994) is possible anymore. In both cases
the magnetic field of the secondary disappears. 

A further hint that the field vanishes when
the secondary becomes a brown dwarf comes from the outburst behaviour of the
dwarf novae in late secular evolution. The modelling of
the extremely bright outbursts and the recurrence time of decades
of WZ Sge stars requires 
an extremely low viscosity in the quiescent accretion disk
(Meyer-Hofmeister et al. 1998). If the accretion disk
viscosity in quiescence is caused by the secondary's magnetic field  
(Meyer \& Meyer-Hofmeister 1999,) this low viscosity value can be
interpreted as due to the disappearance of the secondary star's
magnetic field. 

\subsection{Loss of coupling}
If the magnetic field of the secondary disappears no dipole-dipole
interaction and no conductive tying of primary field to secondary
field are possible. The primary field cannot  
couple to a non-magnetic secondary because buoyancy tends to expell
magnetic fields from the star and there is no convective mixing to
counteract this effect. Ohmic diffusion on the hotter irradiated side
does not suffice for efficient penetration. On the cool surface of
the backside conductivity is too low for magnetic coupling (Meyer \&
Meyer-Hofmeister 1999).

\section{The spin-up phase}
When magnetic coupling is lost the white dwarf starts to
spin up. No angular momentum is returned to the secondary and
thereby to the orbit.

We investigate how the mass transfer develops.
For the structure of the secondary we take a polytrope with pressure
$p$ and density $\rho$ 

\begin{equation}
p = K \rho ^ \gamma ~~~;~~~~~~~~\gamma = 5/3\,.
\end{equation}
$K$ depends on the secondary's mass $M_2$ and radius $R_2$ (Emden
1907). We assume that the star fills its Roche lobe $R_{R,2}$. We use 
Paczy\'nski's (1971) approximation for small mass ratios
$q = M_2/M_1$ ($M_1$ mass of the primary) 

\begin{equation}
\frac{R_{R,2}}{a} = 0.462 \left(\frac{q}{1+q}\right)^{1/3},
\end{equation}
where $a$ is the separation of the binary stars. The separation is
related to the orbital period $P = 2 \pi (a^3/GM)^{1/2},\\  M = M_1 +
M_2,  G$ gravitational constant. The mean density $\overline{\rho}_2$
is then related to  the orbital period $P$,\\ $\overline{\rho}_2 =
110 /(P/\rm h)^2$ (Frank et al. 1985). For a Roche lobe filling
secondary the constant $K$ can be expressed

\begin{equation}
K = 10^{13.08} \left( P_{80} \frac{M_2}{0.07 M_\odot} \right)^{2/3},
\end{equation}
where $P_{80}$ is the orbital period in units of 80 minutes (a
completely degenerate star of cosmic abundance would have K=$10^{12.85}$).

The rate of mass transfer through the Lagrangian  point $L_1$ is given
by the product of density, sound speed and effective cross
section (see Kolb \& Ritter 1990). For $a$ polytrope with $\gamma = 5/3$ we
\,obtain 

\begin{equation}
\dot{M} = \frac{2 \pi a^3 K^{3/2}}{k(q)GM}  
\,8.545\,  \overline{\rho}^2_2 \left[\frac{R_2 - R_{R,2}}{R_2}\right]^3
\end{equation}
with $k(q)$ from the Roche geometry (compare Meyer \&
Meyer-Hofmeister 1983), $k(q)$=5.97 for $q$=0.1. 

For $q = 0.1$ and $M_2 = 0.07 M_\odot$ follows:

\begin{equation}
\dot{M} = 10^{29.93} \frac{M_2/(0.07M_\odot)}{P_{80}} \left[\frac{R_2 -
R_{R,2}}{R_2}\right]^3. 
\end{equation}

We determine the change of $\dot M$. 
We call $\beta$ the fraction of the transferred matter accreted on the
primary, $\dot{M_1}
= \beta \dot{M}$, in the spin-up phase $\beta=1$.

The change of the Roche radius $R_{R,2}$ depends on the change of the
quantities $a$ and $q$.

\begin{equation}
\frac{1}{R_{R,2}} \frac{dR_{R,2}}{dt} = \frac{1}{a}\frac{da}{dt} +
\frac {1}{q(1+q)} \frac{dq}{dt}\,.
\end{equation}

The time derivative of $a$ is related to the derivative of
the angular momentum $J$

\begin{equation}
J = \sqrt{GMa}\frac{M_1M_2}{M}\,.
\end{equation}

We consider orbital angular momentum loss due to gravitational radiation,
spin-up of the white dwarf and, in the propeller phase, expulsion of matter
\begin{eqnarray}
\dot{J} = \dot{J}_{\rm {GW}} + \dot{J}_{\rm {spin-up}}
~~~~~~\rm{spin\!-\!up\,\,phase},
\end{eqnarray}
\begin{eqnarray*}
\dot{J} = \dot{J}_{\rm {GW}} + \dot{J}_{\rm
{propeller}}~~~~~\rm{propeller\,\,phase}.
\end{eqnarray*}

We take $\dot{J}_{\rm {GW}}$ according to Misner et
al. (1973). Additional braking, as considered by Kolb \& Baraffe (1999)
can be taken into account by applying a corresponding factor to 
$\dot{J}_{\rm {GW}}$ and $\dot M_{\rm {GW}}$. $\dot{J}_
{\rm {spin-up}}$ results from the accreted angular momentum used for
spin-up and lost from the orbit

\begin{equation}
\dot{J}_{\rm{ spin-up}} = \sqrt{GM_1a} \left(\frac{r_{\rm{LS}}}
{a}\right)^{1/2}\dot{M},
\end{equation}
where $r_{\rm{LS}}/a= 
\tilde{\omega}_{\rm {d}}$ was evaluated by Lubow $\&$ Shu
(1975), 0.23 for $q$=0.1.

Eqs.(5) to (9) finally yield the equation for the change of $\dot M$.
The quantity $X$ collects all terms proportional to $\dot M$ and is
decisive for the evolution.
\begin{eqnarray}
\frac{d\dot M}{dt}  = C \dot M^{2/3} \left(\dot M_{\rm{GW}}-X\dot M \right),
\end{eqnarray}
\begin{eqnarray*}
C=
\frac{10^{-21.43}}{k(q)^{1/3}} \left(\frac{M_2}{(0.07M_\odot)}\right)^{-2/3}
P_{80}^{-1/3}~~\rm g^{-2/3}\rm s^{-1/3},
\end{eqnarray*}
\begin{eqnarray*}
\dot M_{\rm{GW}}= \frac{10^{14.79}}{q^{2/3}(1-q)^{1/3}}\cdot 
\left(\frac{M_2}{0.07M_\odot}\right)^{8/3}\cdot P_{80}^{-8/3}~
~\rm g\,\rm s^{-1},
\end{eqnarray*}
\begin{eqnarray*}
X &= & \zeta-\frac{1+\beta q}{3(1+q)}- 2f \sqrt{(1+q)(\frac{r_{\rm LS}}{a}})\\
  & & +2(1-\beta q) -(1-\beta)\frac{q}{1+q}\,. 
\end{eqnarray*}

The factor $f$ is the ratio of specific angular momentum lost from the
orbit by the matter flow (either accreted on the white dwarf or
expelled from the system) to the specific angular momentum
carried by the aarriving ccretion stream.
For the spin-up phase $f$ is equal to 1. The mass-radius
exponent $\zeta$=dlnR/dlnM changes from about 0.8 to -1/3 as the
secondaries evolve from a late main sequence star to a (partially)
degenerate cool brown dwarf ( Ritter 1986, Kolb \& Baraffe 1999).

For positive value of $X$ the solution of Eq. (10) tends to the stable
point $\dot M=\dot M_{\rm{GW}}/X$. Due to the change of $X$ with the
start of the
spin-up the mass
transfer rate $\dot M$ increases by more than a factor of ten. We
assume that the matter falling towards the white dwarf in the rotating
magnetosphere behaves diamagnetically and experiences a braking force
from the relative motion between field and fluid (King 1993,
Wynn \& King 1995). The strongest interaction occurs at the point
where the free fall path would reach closest approach and the magnetic
field is largest. If the speed of the rotating magnetic field
becomes faster than that of the material there a new phase involving
expulsion of matter can set in.

For our standard case $M_1=0.7M_\odot$, $M_2=0.07M_\odot$, $P_{\rm
{orbital}}$=80 minutes,  the amount of matter required to reach
critical rotation of the white dwarf is $\Delta M$=0.002$M_\odot$,
accumulated in about $6\,10^6$ys.

\section{The propeller phase}
The falling matter enters the magnetic field of the rotating
dipole. For an inclined dipole the braking and acceleration force
experienced by the matter depends on the rotational phase 
of the dipole. Thus one part of the matter reaches the strong
interaction a bit farther away from the primary and the other comes
closer in. The latter would experience braking and will finally be
accreted. The former can be accelerated outward and flung out of the
system. A propeller effect was already discussed by Illarionov \&
Sunyaev (1975). The interaction of the magnetic field and the matter
stream is complex. We adapt here results obtained from a model for AE
Aquarii by Wynn et al. (1997).

The intermediate polar AE Aqr is observed to spin down at a rate of 
$\dot P_{\rm {spin}}=5.64 \times 10^{-14}$. In the model of Wynn
et al. (1997) the angular momentum given off by the white dwarf adds
to the angular momentum of the stream itself to expell nearly all of
the matter transferred from the secondary. The interpretation of the
observed Doppler tomogram supports this expulsion of matter. Along
these lines we now estimate the orbital angular momentum loss 
$\dot{J}_{\rm{propeller}}$ involved in the expulsion of matter in our
case. We obtain
an estimate for the fraction of matter 1-$\beta$ expelled from the
system by the following consideration. The matter arriving at the
distance of closest approach $r_{\rm{min}}$ needs additional angular
momentum to be accelerated
above escape speed. This is provided by the angular
momentum of the accreted fraction $\beta$.
\begin{eqnarray}
(1-\beta)\left[\sqrt{(1+b^2)} \sqrt{2GM_1 r_{\rm{min}}}
-\sqrt{GM_1r_{\rm{LS}}}\right]
\end{eqnarray}
\begin{eqnarray*}
~~~~~~~= \beta  \sqrt{GM_1r_{\rm{LS}}}.
\end{eqnarray*}
We take for $r_{\rm{min}}$ the value determined by Lubow \&
Shu (1975) $r_{\rm{min}}= a \cdot \tilde{\omega}_{\rm{min}}$. Assuming that the
matter arrives at infinity with
velocity one half of the escape speed at distance $r_{\rm{min}}$ 
from the white dwarf, $b=
v_{\infty}/v_{\rm {escape}}$= 1/2 one obtains
the estimate
\begin{equation}
\beta=1-\frac {1} {\sqrt{2r_{\rm {min}} /r_{\rm {LS}}}}=0.2.
\end{equation} 
The numerical value results for $q$=0.1.

The orbital angular momentum loss in the propeller phase has to be
taken with respect to the binary's center of gravity.
The forces that produce the roughly $90^\circ$ swing around at the primary
before ejection retard the primary's orbital motion and thereby extract
orbital angular momentum. We use the computed
trajectories from Lubow \& Shu, interpolated for $q$=0.1 and write the
estimate for $\dot{J}_{\rm{propeller}}$
\begin{eqnarray}
\dot{J}_{\rm{propeller}} & = & (1-\beta) \dot M \cdot 
\end{eqnarray}
\begin{eqnarray*}                 
~~~~~~~~~~~~~~~     \left[\sqrt{v_{\rm
{escape}}^2+v_{\infty}^2} ~\cdot r_{\rm {min}}+v_{\infty} \frac{q}{1+q}a
\right],
\end{eqnarray*}
where we assume that the expelled matter keeps its angular momentum
with respect to the primary as it rapidly climbs out of the
gravitational potential. The last term on the right side accounts for
the distance between the primary and the binary's center of mass. This yields  
\begin{eqnarray}
\dot{J}_{\rm{propeller}}=\dot M f \sqrt{{GM_1r_{\rm {LS}}}}
\end{eqnarray}
with
\begin{eqnarray}
f=(1-\beta) \sqrt{\frac{2r_{\rm {min}}/a}{r_{\rm {LS}}/a}}
\sqrt{1+b^2} 
\end{eqnarray}
\begin{eqnarray*}
~~~~~~\cdot \left(1+\frac{q}{q+1}\frac{b}{\sqrt{1+b^2}}\frac{1
}{r_{\rm {min}}/a} \right).
\end{eqnarray*}
For $q$=0.1 and $\beta$=0.2 one obtains $f$=1.3. A graphical
evaluation of trajectories calculated by Wynn et al. (1997) for AE Aqr
taking into account the smaller mass ratio leads to $f$=1.5.
We emphasize that these estimates are rough. With such values of $f$
the increase of orbital angular momentum loss compared to
$\dot{J}_{\rm {spin-up}}$ changes the sign of $X$ and then leads to
an accelerated
growth of the mass transfer rate [Eq. (10)]. Whether this occurs
depends on the detailed process during the swing around the white dwarf.
For $f$=1.3 and our standard case the time to reach
arbitrarily large $\dot M$ is only $10^{4.6}$ys.
The rate cannot grow indefinitely, finally the magnetic field becomes
unable
to handle the ever growing mass transfer. The
efficiency of acceleration diminishes and angular momentum loss from
the system gets limited.

\section{Further evolution}
The question arises how such systems develop. If the
system stabilizes at a high transfer rate the secondary may lose all
its mass in a relatively short time and a single fast rotating magnetic
white dwarf remains. The observation of RE J0317-853 (Burleigh et
al. 1999) with these unusual characteristics seems to support such a
possibility.
Another path of evolution might lead to the formation of
an accretion disk, which results in return of angular momentum back to
the secondary, connected initially with a strong increase of accretion
onto the white dwarf, as $\beta$ becomes 1, and then a rapid decrease
of $\dot M$. But the
strong magnetic pressure of the white dwarf removes such a disk before
the mass accretion rate has dropped to the low stable value
$\dot M_{\rm {GW}}/X$ [Eq. (10)]. This could lead to a cyclic
spin-up/spin-down on short timescales, involving phases of high 
$\dot M$ (compare the model for AE Aqr by Wynn et al. 1997) and also
rapid depletion of the secondary. These phases are probably too short
to be observable. If expelled matter would form a circumbinary disk,
such a disk could extract significant angular momentum from the orbit
and thereby even lead to a supersoft X-ray source\,  - \, state (van
Teeseling \& King 1998).

\section{Conclusions}
The loss of magnetic coupling and the degeneracy of the secondary
stars are the important features of the late evolution of AM Her systems.
We conclude that no AM Her systems should exist beyond the period turning
point (except the brown dwarf would be young, hot and still magnetic -
searches for the secondary stars in AM Her systems near period
mimimum would be very desirable).
Instead  many of the unusual systems at low orbital periods might be
descendants of the AM Her systems. There is one parameter, the
strength of the white dwarf magnetic field, which may decide which way
the later evolution goes. 

Dwarf nova systems do not partake in these evolutionary phases. When
they lose magnetic braking and settle to gravitational wave braking
only their mass transfer rate could well become too low for allowing
any outburst (see the marginal case of WZ Sge, Meyer-Hofmeister et
al. 1998) and thus fade from view.

The low viscosity needed to model the outbursts cycles of the dwarf nova
systems in late secular evolution and the disappearance of the AM Her
systems could thus both 
result from the loss of the magnetic field of the companion star, when
the star evolves to a cool brown dwarf. 

\begin{acknowledgements}
We thank  H. Ritter and H.-C. Thomas for helpful discussions. 
\end{acknowledgements}

\end{document}